\documentclass[12pt,a4paper]{article}

\usepackage{authblk}
\usepackage[english]{babel}

\usepackage[letterpaper,top=2cm,bottom=2cm,left=3cm,right=3cm,marginparwidth=1.75cm]{geometry}

\usepackage{amsmath}
\usepackage{amssymb}
\usepackage{graphicx}
\usepackage{natbib}
\usepackage[colorlinks=true, allcolors=blue]{hyperref}
\usepackage{comment}
\usepackage{subcaption}
\usepackage{float}

\newcommand{\til}{~}

\newcommand{\vel}{{\mathbf{v}}}
\newcommand{\br}{{\bar{r}}}
\newcommand{\bq}{{\bar{q}}}
\newcommand{\vr}{{\mathbf{r}}}
\newcommand{\vR}{{\mathbf{R}}}
\newcommand{\vq}{{\mathbf{q}}}

\newcommand{\qpa}{{q_{||}}}
\newcommand{\qpe}{{q_{\perp}}}
\newcommand{\Rpa}{{R_{||}}}
\newcommand{\Rpe}{{R_{\perp}}}

\newcommand{\revise}[1]{{\color{black}#1}}

\title{Density Modulations of Zero Sound}

\author{Leonardo Pisani\thanks{Leonardo.Pisani2@unibo.it}}%
\affil{Dipartimento di Fisica e Astronomia “Augusto Righi”, Università di Bologna, Via Irnerio 46, I-40126, Bologna, Italy \\ INFN, Sezione di Bologna, Viale Berti Pichat 6/2, I-40127, Bologna, Italy}

\begin{document}

\maketitle

\begin{abstract}
We study the density modulation of an interacting Fermi gas caused by the uniform motion of an impurity
at zero temperature. For strong enough interaction among Fermi atoms, the modulation propagates thanks to the excitation of the collective zero sound mode if the impurity speed is above the zero sound threshold. We are able to assess, via a semi-analytic evaluation, the extent of the zero sound contribution to the density oscillation over and above the incoherent background of particle-hole excitations.
Given the strong dependence of the results on the features of the gas interaction potential, we also analyze how they vary 
depending on its \revise{strength, range and shape}.
\end{abstract}

\section{Introduction}

One of the paradigmatic ways of probing the nature of a physical system is that of observing the effects induced by a small object traveling through it. Depending on the system under investigation one observes: 1) ship waves when the body moves on the surface of water \til\cite{Kelvin-1887}, 2) Cherenkov radiation when a charged particle moves through a dielectric medium\til\cite{LL-1984,Carusotto-2006}, 3) shock waves when a source travels at supersonic speed\til\cite{Landau-1987}. 
The density modulation thus produced can vary significantly based on the dispersion relation of the collective modes of the medium excited by the moving object\til\cite{Carusotto-2013}.  

A wake pattern is formed every time the speed of the impurity matches the phase velocity of at least one collective mode. If the excitation spectrum is of the acoustic type, a shock wave is produced \revise{once the source speed becomes} supersonic\til\cite{Landau-1987}. Conversely, on the surface of a deep liquid gravity waves are always found
whatever the speed of the floating object is\til\cite{Carusotto-2013}. 
In the case of an electron liquid the excitation spectrum is predominantly made of  plasma oscillations \revise{
and a shock wave can be excited if the speed of the perturbation is above the speed of sound\til\cite{FETTER-1973,Kolomeisky-2018}. The sound in question has little in common with ordinary hydrodynamic (first) sound, as found in liquids and gases, but rather it resembles a longitudinal wave propagating in an elastic medium, as in a solid\til\cite{Dobbs-2000,GV-2005}.
Owing to the long range nature of the Coulomb repulsion, the frequency of plasma oscillations is much higher than the typical relaxation rates to thermodynamic equilibrium of the electron fluid hence the system remains in a collisionless (non-equilibrium) regime.  
This is in fact reflected in the dispersion law of the plasmon $\omega(k)=\sqrt{\omega_p^2+c_s^2\,k^2}$, where $c_s$ is not the hydrodynamic sound speed of a Fermi gas (which, in first approximation, is $s={v_F\over \sqrt{3}} $  with $v_F$ the Fermi velocity) but its elastic counterpart $s=\sqrt{3 \over 5} v_F$\til\cite{FETTER-1973,GV-2005}. This effect is especially important in the context of plasmonic nanostructures where non-local effects cannot be neglected\til\cite{Du-Liu-2024,Teperik-2013}.

A phenomenon analogous to collisionless plasma oscillations takes place in neutral quantum fluids when finite-ranged and strong interactions are at play, and goes under the name of {\it zero sound}\til\cite{Landau-1957,PN-1966,Pines-1981}.
If the frequency of an exciting perturbation $\omega$ is much higher than the inverse of the hydrodynamic equilibration time $\tau$, the system is said to be in the collisionless regime and responds with the propagation of a longitudinal wave thanks to the presence of strong interactions between its constituents (such role is played by the long-ranged electrostatic field in charged Fermi fluids like plasma). This longitudinal wave is called zero sound and it is generally detected at very low temperatures, where atomic collisions are extremely rare events.
In contrast, at higher temperatures, where the hydrodynamic (collisional) regime holds ($\omega \ll 1/\tau$), ordinary (first) sound is usually found.

The only Fermi liquid where zero sound has been observed so far is $^3$He, originally by means of ultrasound attenuation experiments\til\cite{Abel-1966}, then measuring the dynamic structure factor via inelastic neutron scattering \cite{Scherm-1987,Faak-1994,Glyde-2000} and more recently via inelastic x-ray scattering \cite{Monaco-2007}. In particular, in Ref.\til\cite{Abel-1966} a clear crossover from first to zero sound was observed in the increasing value of the sound speed as the temperature was lowered down to the range of millidegrees Kelvin. It was then demonstrated in terms of a viscoelastic effect that first sound shows viscous behavior (typical of fluids) whereas zero sound manifests its elastic counterpart (typical of solids) and the crossover between the two speeds is epitomized by the relation $c_0^2\simeq c_1^2+{4 \over 3} {G \over \rho}$ , with  $G$ the shear modulus (null in a gas or liquid) and $\rho$ the density\til\cite{Dobbs-2000,Nettleton-1976,Rudnick-1980}. 
As the temperature is lowered causing the collision rate $1/\tau$ to decrease significantly, the response of the system transitions from being liquid-like (with no shear modulus) 
to solid-like with a finite shear modulus originating from the high-frequency regime $\omega \tau \gg 1$\til\cite{Vignale-1999}.

On the theoretical front, zero sound has been initially modeled by means of a generalized Random Phase Approximation (RPA), where a polarization pseudo-potential replaces the bare interaction potential and the effect of both single- and multi-particle-hole ({\it p-h})  excitations is taken into account through a phenomenological expression of the irreducible (screened) polarization function appearing in the RPA density response\til\cite{Pines-1976}. However the RPA failure in reproducing the line-shape of the dynamic structure factor\til\cite{Faak-1994} has prompted the inclusion into the RPA model of a momentum-dependent effective mass thus obtaining a good agreement with experiment\til\cite{Glyde-2000}.
The successive observation in inelastic x-ray scattering of a well defined zero sound mode up to large momentum transfers, where it is supposed to be strongly damped due to the hybridization with the {\it p-h} continuum, has opened a controversy on the actual location of the {\it p-h} band within the full excitation spectrum\til\cite{Schmets-2008,Albergamo-2008}. By taking into account two particle-two hole excitations Krotscheck {\it et al.}\til\cite{Krotscheck-2011} have shown that these pair excitations reduce the damping of the mode in comparison with the RPA response hence contributing to the sharpening of the mode for momenta within the {\it p-h} band.    

The advent of ultracold atomic gases has introduced an exceptionally versatile platform to simulate both quantum matter\til\cite{Bloch-2008,Chin-2010,Baroni-2024,Gross-2017} and quantum many-body theories\til\cite{Giorgini-2008,Bloch-2012}. In particular, density and spin response functions of a Fermi gas have been measured via Bragg spectroscopy \til\cite{Hoinka-2012,Hoinka-2013, Landig-2015,Torma-2016} thus constituting crucial testbeds for quantum many-body calculations.
Recently the renown Lindhard function was measured in a Fermi gas and the microscopic basis of Landau's Fermi liquid theory was investigated\til\cite{Navon-2025}. In the hydrodynamic regime the emergence of first sound was observed in agreement with Landau's transport equation but when the fluid was brought to the collisionless regime no crossover to the zero sound mode was detected. The intrinsically weak and short-ranged nature of the (repulsive) contact interaction in ultracold Fermi gases makes the observability of the zero sound mode very difficult owing to the proximity of its dispersion to the incoherent {\it p-h} band, which in turn induces a strong Landau damping\til\cite{Navon-2025}.

An alternative route to the observation of zero sound was suggested within the context of ultracold dipolar gases\til\cite{Shlyapnikov-2013}.
It was shown that, for a two-dimensional  gas of polar molecules, the coherent peak appearing in the dynamic structure factor and representing the zero sound collective mode proves to be sufficiently separated from the border of the {\it p-h} continuum of incoherent excitations, even when the effective interaction strength between dipoles is weak. Moreover, as the decay rate  of the zero sound turned out to be smaller than that of incoherent {\it p-h} excitations, an actual observation of the mode as a distinct density modulation was also predicted to be possible\til\cite{Shlyapnikov-2013}. 

It is finally worth mentioning that the study of this collective mode appears also in the context of high-density neutron matter, as it provides a much needed physical constraint on the equation of state of nuclear matter and neutron stars\til\cite{Ye-2023}.

}

In this work we consider a neutral Fermi liquid \revise{at zero temperature in a three-dimensional homogeneous geometry} and study the density modulation produced by an impurity moving through it at \revise{constant velocity}. \revise{ The collective mode expected to be excited by the impurity above a certain critical velocity is the zero sound mode}, namely a coherent superposition of {\it p-h} excitations\til\cite{PN-1966,Pines-1981}.
By assuming the impurity to be a weak perturbation respect to the  interaction strength and Fermi energy of the gas, we adopt linear response theory\til\cite{FW-2003,GV-2005} to obtain the density modulation induced by the motion and analyze it by varying the different parameters of the system, that is the magnitude of the impurity velocity, the strength of the gas interaction and its functional form, albeit in a simplified way respect to the \revise{more realistic} paradigm set by $^3$He. 

The paper is organized as follows.
In sec.\til\ref{sec:lrt} we present the linear response theory of the system under study, whereby we adopt the RPA for the treatment of the medium density response, and consider a specific choice of the interaction potential in the medium inspired by the quantum liquid $^3$He.
In sec.\til\ref{sec:zs} we examine the nature of the dispersion relation of zero sound and, based on it, we introduce a polar approximation for the density response function. 
In sec.\til\ref{sec:densmodAN} we take advantage of the polar approximation to obtain a semi-analytic form of the density modulation which allows us to single out the zero sound contribution from that of the incoherent background of {\it p-h} excitations. In sec.\til\ref{sec:densmodNUM} we present our numerical results, compare them with the semi-analytical evaluation of the zero sound contribution and examine how they vary when the different parameters of the system are changed. \revise{ In sec.\til\ref{sec:exprel} possible experimental realizations for the observation of zero sound are suggested.}
Finally, in sec.\til\ref{sec:conc} we draw our conclusions and briefly mention open issues to assess in future work.

\section{Linear response theory}
\label{sec:lrt}

We consider an impurity moving through an interacting Fermi gas with constant velocity $\vel$ and are interested in computing the density modulation induced by it in the medium.
The impurity represents a small perturbation to the system that interacts with itself via finite range repulsive forces. The perturbation is approximated  by the external potential
\begin{equation}
    U_{ext}(\vr,t)= U_{0}\, \delta(\vr-\vel t) 
\end{equation}
whose Fourier transform is
\begin{align}
    \tilde{U}_{ext}(\vq,\omega) &= \int d^3r \, dt \: e^{-i \vq \cdot \vr} e^{i \omega t} \: U_{ext}(\vr,t) \\
     &=  \, \int dt \: e^{i (\omega -\vq \cdot \vel) t}  \: U_0 \\
     &= 2\pi \, \delta(\omega -\vq \cdot \vel) \: U_0.
\end{align}
According to linear response theory \cite{PN-1966,FW-2003,GV-2005}, the density modulation induced by the motion of the impurity has the expression
\begin{align}
    \delta n(\vr,t)&= \int \frac{d^3q}{(2\pi)^3} \frac{d \omega}{(2\pi)} \, e^{i \vq \cdot \vr} e^{-i \omega t} \, \chi(\vq,\,\omega+i\,0^+)  \tilde{U}_{ext}(\vq,\omega)  \label{eq:ftchi} \\
    &= \int \frac{d^3q}{(2\pi)^3} \, e^{i \vq \cdot (\vr-\vel t)}\, \chi(\vq,\,\vq \cdot \vel+i\,0^+) \, U_{0}, \label{equ:dn}
\end{align}
where $\chi(\vq,\omega+i\,0^+)$ is the Fourier transform of the retarded
density response function (or density susceptibility).
Therefore, once the strength of the external potential is fixed, the computation of the induced density modulation boils down to that of the spatial Fourier transform of $\chi(\vq,\vq \cdot \vel + i\,0^+)$.

\revise{
In analogy to the electron gas responding to a point-like test-charge, the density response function represents the polarization of the fermionic medium as a result of two simultaneous agents: the external potential and the potential self-consistently generated by the induced density modulation. 
The resulting effective potential reads in Fourier space,  
\begin{align}
\tilde{U}_{eff}(\vq,\omega)&=\tilde{U}_{ext}(\vq,\omega) + \tilde{V}(\vq)\, \delta \tilde{n}(\vq,\omega)   \nonumber \\
&=\tilde{U}_{ext}(\vq,\omega) + \tilde{V}(\vq)\, \chi(\vq,\omega+i\,0^+) \, \tilde{U}_{ext}(\vq,\omega)  \nonumber \\
&=\tilde{U}_{ext}(\vq,\omega) \left[ 1 + \tilde{V}(\vq)\, \chi(\vq,\omega+i\,0^+) \right]  
\label{eq:ueffext}
\end{align}
where $\tilde{V}(\vq)$ is the Fourier transform of the interaction potential in the medium and $\delta \tilde{n}(\vq,\omega)$ is the Fourier transform  defined implicitly in Eq.\til\eqref{eq:ftchi}. We note that Eq.\til\eqref{eq:ueffext}  contains the definition of the generalized dielectric function
$\epsilon(\vq,\omega)^{-1}=1+ \tilde{V}(\vq)\, \chi(\vq,\omega)$.

Alternatively, one can introduce a density response (o polarization) function $\bar\chi(\vq,\omega)$ originating from the total effective potential $\tilde{U}_{eff},(\vq,\omega)$
\begin{equation}
    \delta n(\vr,t)= \int \frac{d^3q}{(2\pi)^3} \frac{d \omega}{(2\pi)} \, e^{i \vq \cdot \vr} e^{-i \omega t} \, \bar\chi(\vq,\,\omega+i\,0^+)  \tilde{U}_{eff}(\vq,\omega),
\label{eq:Ueff}
\end{equation}
thus obtaining the relation  
\begin{align}
\tilde{U}_{eff}(\vq,\omega)
&=\tilde{U}_{ext}(\vq,\omega) + \tilde{V}(\vq)\, \bar\chi(\vq,\omega+i\,0^+) \, \tilde{U}_{eff}(\vq,\omega). 
\label{eq:ueffext2}
\end{align}
The special case where $\tilde{U}_{ext}(\vq,\omega) \equiv V(\vq)$ allows us to recognize (without loss of generality) that the quantity $\bar\chi(\vq,\omega+i\,0^+)$ is the proper or irreducible density response function, namely it includes all virtual (diagrammatic) processes which cannot be separated in two parts by cutting a single potential line\til\cite{FETTER-1973}. 
As it is a diagrammatically involved quantity to compute,
in this work we consider the simplest possible approach of approximating $\bar\chi(\vq,\omega+i\,0^+)$ with the density response function of a non-interacting Fermi gas $\chi_0(\vq,\omega+i\,0^+)$. This approach was pioneered by Bohm and Pines\til\cite{Bohm-1951}
in the quantum treatment of the collective excitations of the electron gas and given the name of RPA. It is equivalent to treating the fermion-fermion interaction at a dynamical mean-field level (time-dependent Hartee theory) and to neglecting all remaining effects due to correlation and exchange embedded in $\bar\chi(\vq,\omega)$.

By substitution of Eq.\til\eqref{eq:ueffext} into Eq.\til\eqref{eq:Ueff} and comparison with Eq.\til\eqref{eq:ftchi}, one obtains the following relation between the two response functions,
\begin{equation}
\chi(\vq,\omega+i\,0^+)=\bar\chi(\vq,\omega+i\,0^+) +\bar\chi(\vq,\omega+i\,0^+) \, \tilde{V}(\vq)\,  \, \chi(\vq,\omega+i\,0^+),
\label{eq:rpa}
\end{equation}
which can then be cast in the following familiar form  \cite{PN-1966,FW-2003,GV-2005}
\begin{align}
    \chi(\vq,\omega+i\,0^+) &= \frac{\bar\chi(\vq,\omega+i\,0^+)}{1-\tilde{V}(\vq)\bar\chi(\vq,\omega+i\,0^+)} \\
    &\approx \frac{\chi_0(\vq,\omega+i\,0^+)}{1-\tilde{V}(\vq)\chi_0(\vq,\omega+i\,0^+)},
\end{align}
where $\chi_0(\vq,\omega)$ is the density response function of the corresponding non-interacting Fermi gas. 
} The latter can be computed analytically and reads
\begin{equation}
\chi_0(\vq,\omega+i\,0^+)=- N_0 \; f_L(z,u),
\end{equation}
with $N_0=mk_F/\pi^2$, the total density of states per unit volume at the Fermi level  and $f_L(z,u)$ the Lindhard function \cite{FW-2003,GV-2005}
\begin{equation}
f_L(z,u)=\frac{1}{2}+\frac{1-(z-u)^2}{8 z} \log \frac{z-u+1}{z-u-1}+\frac{1-(z+u)^2}{8 z} \log \frac{z+u+1}{z+u-1},
\end{equation}
with 
$u=\frac{\omega+i\,0^+}{v_F q}$ and $z=\frac{q}{2 k_F}$ with $q=|\vq|$.

\begin{comment}
\begin{align}
\frac{\Re[\chi_0(\vq,\omega)]}{N(0)} & = \frac{1}{4
   q}\left[1-\left(\frac{\omega}{2 q}-\frac{q}{2}\right)^2\right] \log
   \left(\frac{\frac{\omega}{2 q}-\frac{q}{2}+1}{-\frac{\omega}{2 q}+\frac{q}{2}+1}\right) \nonumber \\
   & -\frac{1}{4
   q}\left[1-\left(\frac{\omega}{2 q}+\frac{q}{2}\right)^2\right] \log
   \left(\frac{\frac{\omega}{2 q}+\frac{q}{2}+1}{-\frac{\omega}{2 q}-\frac{q}{2}+1}\right)-\frac{1}{2},
\end{align}
\begin{align}
\frac{\Im[\chi_0(\vq,\omega)]}{N(0)} & =\frac{\pi}{4} \text{sgn}(\omega) 
\begin{cases}
 \hfil 0 & | \omega | >q^2+2 q \; \text{or} \; \left(q>2 \;  \& \; \left| \omega\right| <q^2-2 q \right), \\
 \hfil \frac{|\omega| }{q} & q<2 \; \& \; |\omega| <| q^2-2q|,  \\
 \frac{1}{q}\left| 1-\left(\frac{|\omega| }{2
   q}-\frac{q}{2}\right)^2\right|  & q^2+2q
   >|\omega| >|q^2-2q|,
\end{cases}
\end{align}    
\end{comment}

Concerning the Fermi gas interaction potential $\tilde{V}(\vq)$, \revise{we consider only finite-range forces and adopt a simple form of the interaction potential directly in Fourier space,
\begin{equation}
\tilde{V}\left(\frac{q}{k_F}\right)= \frac{\tilde{V}_0}{N_0} e^{-\left( \frac{q}{k_F} \, \times \,k_F r_0\right)^\alpha}, \quad \quad \tilde{V}_0=\frac{2}{\pi}\frac{V_0}{\epsilon_F},
\end{equation}
where $V_0$ is the strength of the interaction in real space, $r_0$ plays the role of the range of the interaction in real space and $\alpha$  simulate the sharpness of the potential in momentum space. Assuming the interaction in real space to be of the form $V(r)=V_0\,v(r)\, \gamma$, with $\gamma$ to be determined by the condition $V(r=0)=V_0$ and $v(r)$ a well-behaved and decaying function of $r$,  the relation between $V(r)$ and $\tilde{V}(q)$ is given by,
\begin{align}
    \tilde{V}\left(\frac{q}{k_F}\right)&=  \int d^3r \, e^{-i \, \vq \cdot \vr}\, V(r)  \nonumber \\
        &=4\pi \int_0^{+\infty} dr\,r^2\,V(r) \, \frac{\sin qr}{qr} \nonumber \\
\{r\to \br=k_Fr, \, q\to \bq=q/k_F \}        
        &= 4\pi \, \frac{V_0}{k_F^3} \, \gamma \int_0^{+\infty} d\br\,\br^2\,v\left(\frac{\br}{k_F}\right) \, \frac{\sin \bq \br}{\bq \br} \nonumber \\
        &=\frac{2}{\pi}\, \frac{V_0}{\epsilon_F} \frac{\pi^2}{m k_F} \, \gamma \int_0^{+\infty} d\br\,\br^2\,v\left(\frac{\br}{k_F}\right) \, \frac{\sin \bq \br}{\bq \br} \nonumber \\
        &=\frac{\tilde{V}_0}{N_0} \, \gamma\int_0^{+\infty} d\br\,\br^2\,v\left(\frac{\br}{k_F}\right) \, \frac{\sin \bq \br}{\bq \br}, \nonumber 
\end{align}
which implicitly defines $v(r)$ as
\begin{equation}
    e^{-\left( \bq \, \times \,k_F r_0\right)^\alpha}= \gamma \int_0^{+\infty} d\br\,\br^2\,v\left(\frac{\br}{k_F}\right) \, \frac{\sin \bq \br}{\bq \br},
\end{equation}
namely by inverse Fourier transform 
\begin{equation}
\gamma \, v(r) = \frac{4\pi}{k_F^3} \int \frac{d^3q}{(2\pi)^3} \, e^{i\, \vq \cdot \vr} \, e^{-\left( q \, \times r_0\right)^\alpha},
\end{equation}
and as a result
\begin{equation}
    \gamma^{-1}=v(0)=\frac{2 \,\Gamma \left(\frac{3}{\alpha}\right)}{\pi\,  \alpha \, (k_Fr_0)^3}.
\end{equation}
}

Throughout this work we consider the representative choice of gas interaction parameters $\tilde{V}_0=4,\,k_Fr_0=1,\,\alpha=8$,
and then examine the effect of changing their values.
This choice is inspired by the characteristics of the effective interaction at play in $^3$He\til\cite{Aldrich-1978,Glyde-2000,Krotscheck-2010}, where the strength $V_0$ is about 4-5 times the Fermi energy $\epsilon_F$ \revise{and the decay of the interaction potential in momentum space is close to $\alpha=8$\til\cite{Krotscheck-2010}}.

The strength of the impurity potential 
is given by the dimensionless quantity $\tilde{U}_0 = N_0 U_0$ and is taken to be a weak perturbation. Since it must be much smaller than the gas interaction strength $\tilde{V}_0$, we fix its value at $\tilde{U}_0=0.1$ throughout this work.

Finally we  recast the expression of the density modulation\til(\ref{equ:dn}) in the following explicit form
\begin{equation}
    \delta n(\vr,t)=- \int \frac{d^3q}{(2\pi)^3} \, e^{i \vq \cdot (\vr-\vel t)}\, \frac{ f_L\left(\frac{q}{2 k_F},\frac{\vq \cdot \vel+i\,0^+}{v_F q} \right) }{ 1+\tilde{V}_0 e^{-(q \,r_0)^\alpha} \; f_L\left(\frac{q}{2 k_F},\frac{\vq \cdot \vel++i\,0^+}{v_F q} \right) } \; \tilde{U}_0, \label{equ:dn2}
\end{equation}
where $\tilde{V}_0$ and $\tilde{U}_0$ represent the dimensionless strength of the gas interaction and of the perturbation, respectively.

\section{Zero sound}
\label{sec:zs}

\begin{figure}[ht]
\centering
\includegraphics[width=0.5\textwidth]{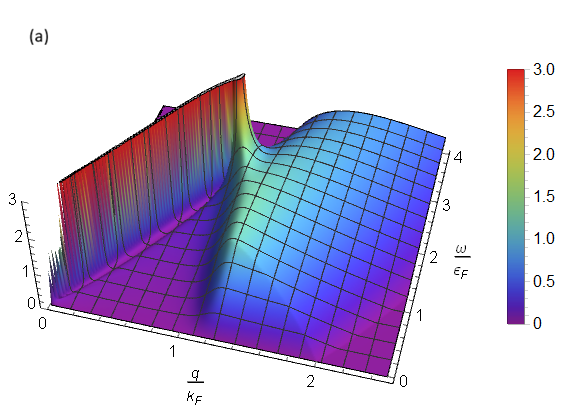}
\includegraphics[width=0.475\textwidth]{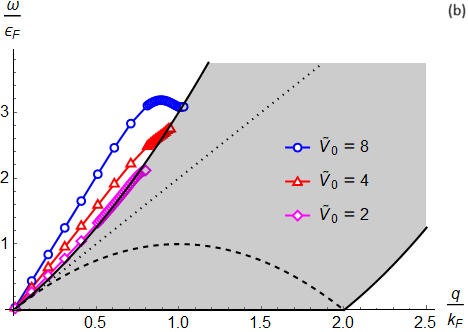}
\caption{\label{zs0} (a) Dynamic structure form factor at $\omega>0$ showing the continuum of incoherent {\it p-h} excitations and the undamped collective excitation of the zero sound mode (sharp peaks). Gas parameters: $\tilde{V}_0=4,\, k_Fr_0=1,\,\alpha=8.$
(b) Zero sound dispersion relation for increasing values of $\tilde{V}_0=2,4,8$ (diamonds, triangles and circles respectively) and {\it p-h} continuum (gray region). Dotted line $\omega=v_F q$ in normalized units. The lower half of the spectrum is omitted for symmetry reasons. }
\end{figure}

The imaginary part of $\chi(\vq,\omega+i\,0^+)$ is proportional to the structure form factor, which provides the spectrum of density excitations\til\cite{PN-1966,FW-2003}. The quantity
$\Im \chi(\vq,\omega+i\,0^+)/N_0$ is reported in Fig.\ref{zs0}(a), whereby, in addition to a continuum of incoherent {\it p-h}  excitations, the sharp peaks (artificially broadened for illustrative purposes) provide the dispersion of the undamped collective mode, that is the zero sound \cite{PN-1966}. 

In Fig.\til\ref{zs0}(b) the zero sound dispersion \revise{$\Omega(q)$} is shown for increasing values of the gas interaction $\tilde{V}_0=2,4,8$. 
The slope of the dotted line is the Fermi velocity $v_F$. The speed of zero sound at long wavelengths $c_0=\lim_{q\rightarrow 0}\revise{\frac{\Omega(q)}{q}}$ is strictly larger than the Fermi velocity\til\cite{FW-2003} and increases with the interaction strength.
It is evident that for weak interaction strength the mode is very close to the continuum threshold, hence more subject to decay into {\it p-h} pairs, whereas at larger strength the mode is well separated from the {\it p-h} band and expected to be dominant over said band.
Contrary to ordinary sound, zero sound does not exist for all wavelengths
but rather it disappears by entering the continuum and becoming (Landau) damped.
The momentum at which it disappears is determined by both the strength of the interaction and the way the latter decays with momentum.

An impurity can act as a probe of this collective excitation if its velocity $\vel$ is larger than the zero sound velocity $c_0$.
In the present study we consider only the regime of strong interaction of the fermionic medium $(V_0 > \epsilon_F)$. The intensity plot in Fig.\til\ref{zs0}(a) clearly shows that in this regime the dominant contribution to the density excitations spectrum is represented by the zero sound mode with respect to the incoherent {\it p-h} continuum. This allows us to adopt the following polar form for the density response function
\begin{equation}
    \chi(q,\omega+i\,0^+)= W(q,\Omega(q))\left[\frac{1}{\omega+i\,0^+ -\Omega(q)}-\frac{1}{\omega+i\,0^+ +\Omega(q)} \right],
\end{equation}
where $\Omega(q)$ is the positive branch of the zero sound dispersion shown in Fig.\til\ref{zs0}(b) and $W(q,\Omega(q))$ its spectral weight,
\begin{equation}
    W(q,\Omega(q)) = \frac{\chi_0(q,\Omega(q))}  {\left| \frac{\partial}{\partial \omega} 
    \left[ 1-\tilde{V}(q)\chi_0(q,\omega+i\,0^+)  \right] \right|}_{\omega=\Omega(q)} .
\end{equation}
Both $\Omega(q)$ and $W(q,\Omega(q))$ are evaluated numerically and then fitted to a cubic spline. In this way we are able to obtain a semi-analytical expression for Eq.\til(\ref{equ:dn}), which not only represents a benchmark for our numerical calculations but also singles out the contribution of zero sound to the density modulation from that of the incoherent band. This will be presented in the next section.

\section{Induced density modulation}

\subsection{Semi-analytical results}
\label{sec:densmodAN}

In the following we illustrate an approximate and semi-analytic evaluation 
of Eq.\til(\ref{equ:dn}), valid when $v=|\vel|>c_0$. 
Given the geometry of the problem, we introduce cylindrical coordinates
$(q_{||},q_\perp)$ with $q_{||}$ along  the polar axis $\vel$ and $q_{\perp}$ perpendicular to it, 
and adopt the shorthand notation $\vR=\vr-\vel\,t$ \cite{FETTER-1973}.
Eq.\til(\ref{equ:dn}) thus takes the expanded form
\begin{align}
    \delta n(\vR)&= \int_0^{+\infty} \frac{\qpe d \qpe}{2\pi} \, \int_{-\infty}^{+\infty} \frac{d \qpa}{2\pi} \, \int_{0}^{2\pi} \frac{d \theta}{2\pi} \; e^{i \,\qpa  \Rpa + i \, \qpe  \Rpe \cos\theta}\, \chi\left(\sqrt{\qpa^2+\qpe^2},\, \qpa \, v +i\,0^+ \right) \, U_{0}, \\
    &= \int_0^{+\infty} \frac{\qpe d \qpe}{2\pi} \, J_0( \qpe  \Rpe) \, \int_{-\infty}^{+\infty} \frac{d \qpa}{2\pi} \; e^{i \,\qpa  \Rpa}\, \chi\left(\sqrt{\qpa^2+\qpe^2},\, \qpa \, v +i\,0^+ \right) \, U_{0}. \label{equ:dncil}
\end{align}
where $J_0(x)$ is the zero order Bessel function of first kind, the dependence of $\chi$ on $|\vq|$ rather than  $\vq$ has been made explicit and $\Rpa=r_{||}-|v|t, \; \Rpe=r_\perp$ are the spatial cylindrical coordinates in the impurity reference frame with $\Rpa$ ($r_{||}$) along the polar axis $\vel$. In the following the dependence of $\Rpa$ on $t$ is kept implicit in the short hand notation ad implies that $\Rpa>0$ is the region ahead of the impurity whereas $\Rpa<0$ is that behind it.

\begin{figure}[ht]
\begin{subfigure}[h]{0.35\columnwidth}
\includegraphics[width=\linewidth]{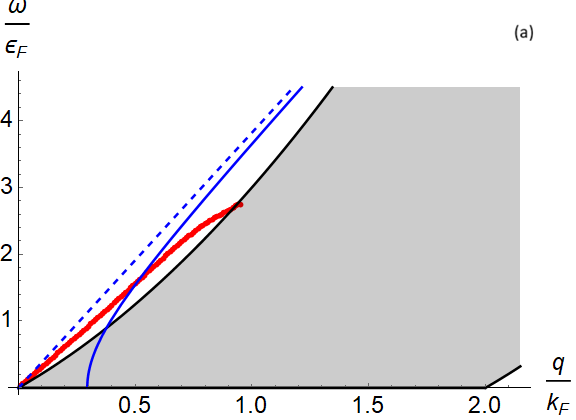}
\end{subfigure}
\begin{subfigure}[h]{0.3\columnwidth}
\includegraphics[width=\linewidth]{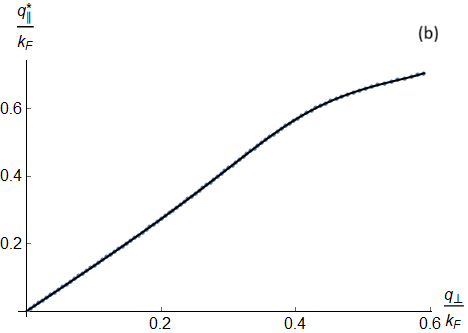}
\end{subfigure}
\begin{subfigure}[h]{0.32\columnwidth}
\includegraphics[width=\linewidth]{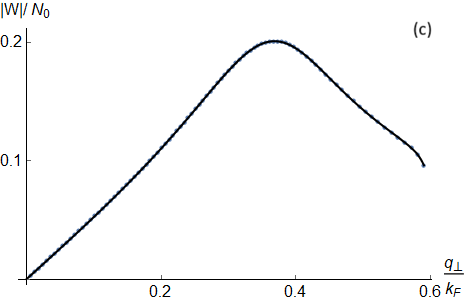}
\end{subfigure}
\caption{\label{zsqpath} (a) Integration path of the cylindrical coordinate $q_{||}$ on the $(q,\omega)$ plane of Fig.\til\ref{zs0}(b).
(b) Dependence of the pole $q_{||}^*$ on $q_\perp$. (c) Spectral weight $W$ of the pole $q_{||}^*$ as a function of $q_\perp$.
Parameters: $\tilde{V}_0=4,\, k_Fr_0=1,\, \qpe=0.3k_F,\, v=1.25 c_0$.}
\end{figure}

The integration on $\qpa$ in Eq.\til(\ref{equ:dncil})  can be represented on the $(q,\omega)$ plane of Fig.\til\ref{zs0}(b)   by drawing, for each choice of $\qpe$, the hyperbola of parametric equations 
\begin{subequations}
\label{equ:paramhyp}
\begin{align}
q&=\sqrt{\qpa^2+\qpe^2}, \\
\omega&=\qpa \, v,    
\end{align}
\end{subequations}
as illustrated in Fig.\til\ref{zsqpath}(a) with a blue solid line for the specific choice, $\qpe=0.3 k_F$ and $v=1.25 c_0$.

As argued previously, we adopt the approximation that the dominant contribution to the density susceptibility is provided by the zero sound pole. This allows the evaluation of the integral on $q_{||}$ in Eq.\til(\ref{equ:dncil}) analytically by extending it to the complex domain $z_{q_{||}}$. 
On this plane, the original two poles $z_\pm=\pm\Omega(q)-i\,0^+$ of $\chi(q,z)$ are remapped according to the  equations  $\pm \Omega\left(\sqrt{(z^*_{q_{||}})^2+\qpe^2}\right)-i\,0^+= z^*_{q_{||}}v$ 
for a given $q_\perp$. In the case of small $q_\perp$ one can approximate $\Omega(q)\simeq c_0 q$ and obtain an analytical expression for the poles
\begin{equation}
z^*_{q_{||}} \simeq \pm\frac{c_0}{\sqrt{v^2-c_0^2}}q_\perp-i \, 0^+.    
\end{equation}
It is crucial to note that the two poles are always positioned in the lower half $\Im{z_{q_{||}}}<0$ owing to the positive infinitesimal $i\,0^+$\til\cite{FETTER-1973}.

\begin{figure}[ht]
\centering
\includegraphics[width=0.4\textwidth]{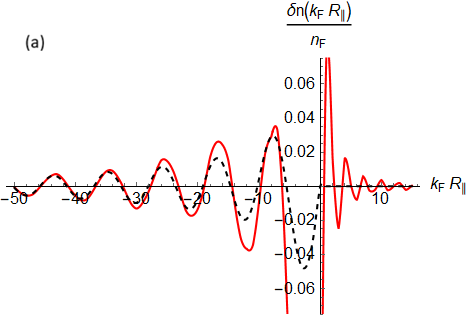}
\includegraphics[width=0.55\textwidth]{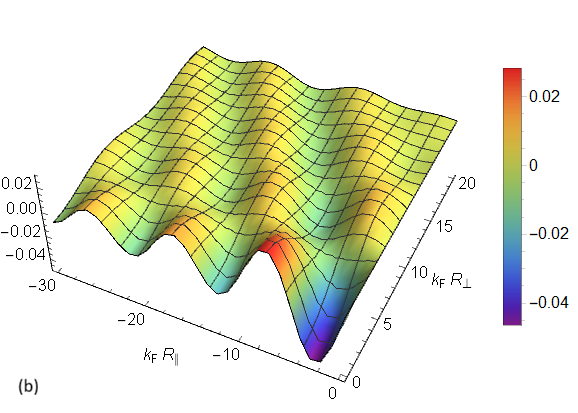}
\caption{ (a) Density modulation induced by a mobile impurity  along its line of motion $(\Rpe=0)$ with velocity $v=1.25c_0$ and  perturbing potential strength  $\tilde{U}_0=0.1$.
The dashed line reports the zero sound contribution given by  Eq.\til(\ref{equ:dnanal}). 
Gas interaction parameters: $\tilde{V}_0=4,\, k_Fr_0=1,\,\alpha=8$. (b) Same as (a) but including the  dimension perpendicular to the line of motion $(\Rpe > 0)$. }
\label{drho0}
\end{figure}

Therefore for $R_{||}>0$, that is ahead of the impurity, one must close the integration contour on the upper half plane, ending up with  the vanishing of  Eq.\til(\ref{equ:dncil}). On the other hand, if $R_{||}<0$, that is behind the impurity, both poles are  included in the integration contour and one obtains
\begin{equation}
    \int_{-\infty}^{+\infty} \frac{d \qpa}{2\pi} \; e^{i \,\qpa  \Rpa}\, \chi\left(\sqrt{\qpa^2+\qpe^2},\, \qpa \, v +i\,0^+ \right)= -2\; \Theta(-R_{||})\; \sin{\left(q_{||}^{*} \, R_{||}\right)}  \; \frac{W\left(q^{*},\Omega(q^{*})\right)}{v},
\end{equation}
where $q_{||}^*\equiv\Re z^*_{q_{||}}$ is a function of $\qpe$ and $q^*=\sqrt{{q^*_{||}}^2+\qpe^2}$.
Fig.\til\ref{zsqpath}(b) and (c) illustrate the behavior of the pole and its spectral weight respectively, for a representative choice of the gas parameters.

The remaining integration on $q_\perp$ is then performed numerically with an upper cutoff dictated by the momentum $q_\perp^c$ at which the zero sound mode enters the continuum band, thus yielding the final expression
\begin{equation}
    \delta n_\mathrm{zs}(\vr,t)  = \,-2 \, \Theta(-R_{||}) \int_0^{q_\perp^c} \frac{\qpe d \qpe}{2\pi} \, J_0( \qpe  \Rpe)  \; \sin{\left(q_{||}^{*} \, R_{||}\right)}  \; \frac{W\left(q^{*},\Omega(q^{*})\right)}{v} \, U_{0}. \label{equ:dnanal}
\end{equation}
In addition to representing a robust and controlled semi-analytic benchmark of our numerical calculations, expression\til(\ref{equ:dnanal}) allows us to isolate the contribution of the zero sound mode to the density modulation from that stemming from the incoherent {\it p-h} excitation band.

\subsection{Numerical results}
\label{sec:densmodNUM}

The full numerical evaluation
of the Fourier transform in  Eq.\til(\ref{equ:dn}) is performed by means of the software  Wolfram Mathematica \cite{Mathematica}.

\begin{figure}[ht]
\centering
\includegraphics[width=0.7\textwidth]{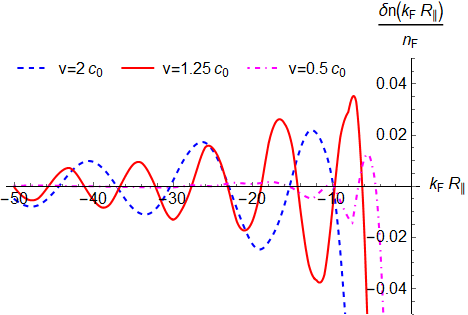}
\caption{\label{drhovel} Density modulation for varying impurity velocity from below to above the zero sound threshold $c_0$. 
Gas parameters: $\tilde{V}_0=4,\, k_Fr_0=1,\,\alpha=8$.}
\end{figure}

In Fig.\til\ref{drho0} we show the density modulation computed 
for an impurity moving through the gas at velocity $v=1.25c_0$ and perturbing it with the strength $\tilde{U}_0=0.1$. The gas interaction parameters  are fixed at the values $\tilde{V}_0=4,\,k_Fr_0=1,\,\alpha=8$. 
In Fig.\til\ref{drho0}(a) we first consider the density modulation along the line of motion $R_\perp=0$ (solid line), where the induced effect is larger: we notice that beyond the immediate vicinity of the impurity ($k_F R_{||} < - 5$) the contribution of the zero sound motion to the density modulation is dominant over that of the incoherent {\it p-h} background, as
the benchmark of Eq.\til(\ref{equ:dnanal}) (black dashed line) is gradually and fully  reached. 

Being mainly interested in the behavior of the modulation far from the impurity where the approximation (\ref{equ:dnanal}) is found to capture the full density modulation, in Fig.\til\ref{drho0}(b) we present a plot of the density variation along both $R_{||} $ and $R_{\perp}$ generated by Eq.\til(\ref{equ:dnanal}). 
The density modulation extends across the whole space, in stark contrast to the case of plasma oscillations of the electron gas or gravity waves on the surface of a liquid whereby the Mach cone $\sin\theta=c/v$ defines the region of propagation of the density modulations\til\cite{Kolomeisky-2018,Carusotto-2013}. 
However we notice that the wave fronts in Fig.\til\ref{drho0}(b) 
propagate at an angle $\tan\theta=R_\perp/R_{||}=v/c$,
with $(R_{||},R_\perp)$ the coordinates of a point on the crest or trough of a wave front.
The fundamental difference with the aforementioned systems is found in the limited momentum range of existence of the zero sound dispersion.

It is also very interesting to analyze the density response of the medium as the velocity of the impurity changes from below the zero sound threshold,  say $v=0.5c_0$ to above it, say $v=2c_0$. This evolution is  reported in Fig.\til\ref{drhovel}. At subsonic speed we notice that the density response is drastically damped and practically disappears for $k_F R_{||} < - 20$. As the velocity increases, the range of density modulation expands and  when the velocity exceeds that of the zero sound, a remarkable increase in the strength and range of the  response takes place. 
We thus find clear evidence of long range effects by which one is able to discriminate the two regimes above and below $c_0$. This effect increases with the velocity of the impurity.

\begin{figure}[ht]
\centering
\includegraphics[width=0.475\textwidth]{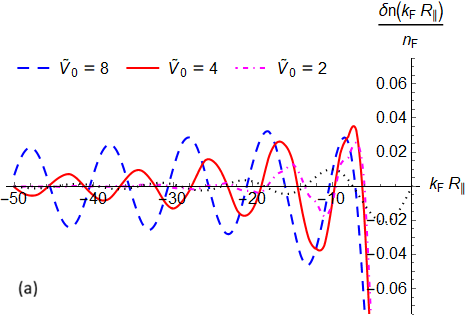}
\includegraphics[width=0.475\textwidth]{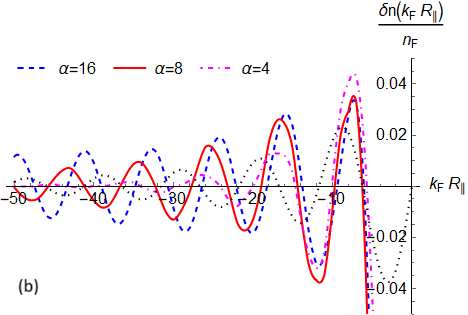}
\caption{(a) Density modulation for varying strength $\tilde{V}_0$ of the gas interaction. The black dotted line corresponds to Eq.\til(\ref{equ:dnanal}) for $\tilde{V}_0=2$. (b) Density modulation for varying sharpness $\alpha$ of the gas potential $\tilde{V}(q)$. The black dotted line corresponds to Eq.\til(\ref{equ:dnanal}) for $\alpha=4$.
In both panels all other parameters are as in Fig.\til\ref{drho0}.}
\label{drhoV0alpha}
\end{figure}

We now focus on the regime $v=1.25c_0$ of Fig.\til\ref{drho0} and examine how these results change when varying the interaction parameters, by computing the density modulation along the line of motion, where the effect of the perturbation is larger. In Fig.\til\ref{drhoV0alpha}(a) we show the evolution of $\delta n$
as the gas interaction strength is varied from $\tilde{V}_0=2$ to $\tilde{V}_0=8$, keeping all remaining parameters as in Fig.\til\ref{drho0}.
We find a strong dependence on $\tilde{V}_0$ not only of the full density response but also of the zero sound contribution. For the smallest coupling $\tilde{V}_0=2$ we also report
the approximated expression Eq.\til(\ref{equ:dnanal}) with a dotted black line as a representative case where this approximation breaks down. This can be ascribed to the characteristics
of the zero sound dispersion found in Fig.\ref{zs0}(b): for the lower coupling 
the dispersion is extremely close to the continuum threshold thus experiencing  stronger damping than in the other two cases. As a consequence the polar approximation adopted to obtain Eq.\til(\ref{equ:dnanal}) ceases to be valid.
\begin{figure}[H]
\centering
\includegraphics[width=0.65\textwidth]{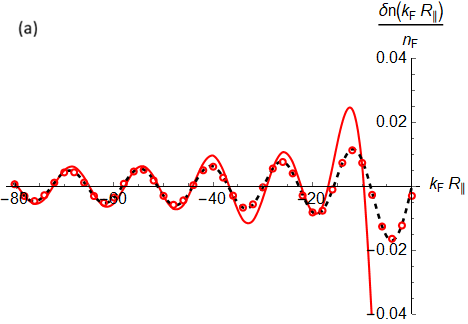}
\includegraphics[width=0.65\textwidth]{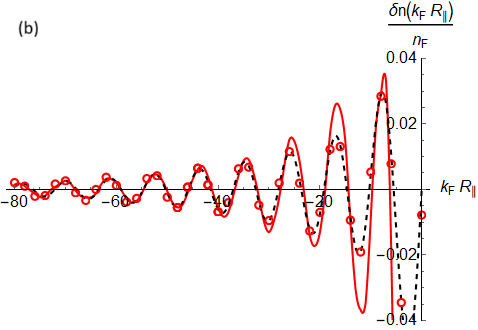}
\includegraphics[width=0.65\textwidth]{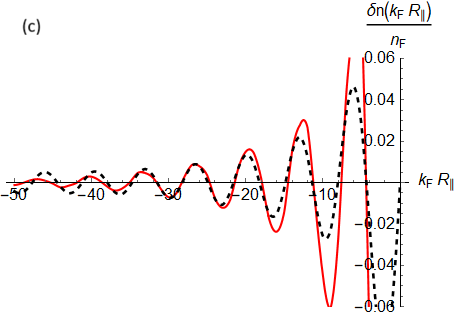}
\caption{Density modulation for $\tilde{V}_0=4$ and $v=1.25c_0$ shown as a red solid line for three different values of the interaction range, (a) $k_Fr_0=1.5$, (b)  $k_Fr_0=1.0$ and (c) $k_Fr_0=0.8$. The black dashed line represents the zero-sound contribution given by Eq.\til\eqref{equ:dnanal}. 
In panels (a) and (b) the zero-sound contribution stemming from the numerical evaluation of Eq.\til\eqref{equ:dn2} for momenta $0 \leq q_\perp \leq q_\perp^c$, is also reported as red circles.}
\label{drhokFr0}
\end{figure}

We also vary the exponent $\alpha$ to simulate the sharpness of the interaction potential in momentum space \revise{(see Fig.\til\ref{drhoV0alpha}(b))}. This feature is relevant because it impacts the damping of the mode when it touches the continuum threshold and if the damping interests an extended range of momenta then the approximation\til(\ref{equ:dnanal}) is not valid any more. This is the case for the $\alpha=4$ curve (dash-dotted line) which is seen to depart from the zero sound benchmark (dotted black line) in contrast to the other two values of $\alpha$ (benchmark not shown ).

Finally we assess the dependence of the results on the interplay between the range $k_Fr_0$ and the strength $\tilde{V_0}$ of the interaction. 
In Fig.\til\ref{drhokFr0} the density modulation at $\tilde{V}_0=4$ and $v=1.25c_0$ is shown for three different values of the interaction range: (a) $k_Fr_0=1.5$, (b)  $k_Fr_0=1.0$ and (c) $k_Fr_0=0.8$ as a red solid line. The black dashed line represents the semi-analytical estimation of the zero-sound contribution given by Eq.\til\eqref{equ:dnanal}. 
We also show in panels (a) and (b) the zero-sound contribution stemming from the evaluation of Eq.\til\eqref{equ:dn2} for momenta $0 \leq q_\perp \leq q_\perp^c$ as red circles, thus representing the numerical counterpart of Eq.\til\eqref{equ:dnanal} (black dashed line), with which it is found to agree optimally.
For these two larger ranges (panels (a) and (b)) we see that at long distances the only contribution to the density modulation (solid red line) is that of zero sound (dashed line and circles), 
whereas for the smaller range (panel (c)) this contribution is lost at large distances, signaling the breakdown of Eq.\til\eqref{equ:dnanal} and a significant damping of the zero sound mode.

\begin{figure}[H]
\centering
\includegraphics[width=0.495\textwidth]{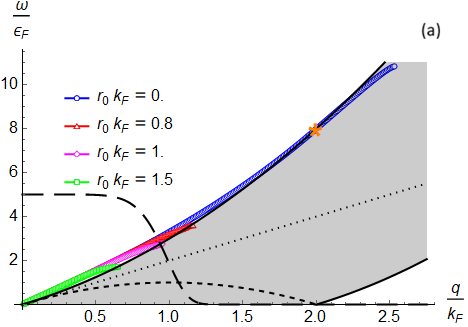}
\includegraphics[width=0.495\textwidth]{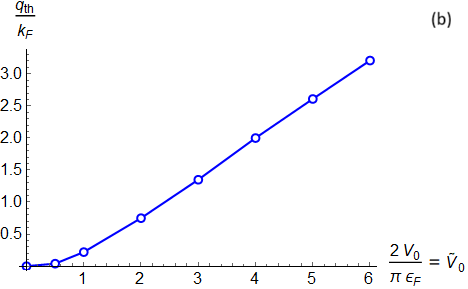}
\caption{(a) Dispersion relations of zero sound for range values corresponding to Fig.\til\ref{drhokFr0} along with the case of a zero-range (contact) potential. The long-dashed line illustrates the interaction potential $\tilde{V}(\vq)$ for $k_Fr_0=1$. 
The orange star identifies the momentum $q_\mathrm{th}$ and energy at which the dispersion enters the continuum band in the case of a vanishing range potential. (b) Correspondence between the threshold momentum $q_\mathrm{th}$ and the interaction strength $\tilde{V}_0$.}
\label{dispkFr0V0qc}
\end{figure}

To clarify the rationale behind this behavior, we show in Fig.\til\ref{dispkFr0V0qc}(a) the zero sound dispersion curves for the corresponding interaction ranges 
along with the dispersion curve for a negligible value of the range $(k_F r_0 \simeq 0)$ in order to illustrate the intrinsic nature of the zero sound mode: it is seen to enter the {\it p-h} continuum  at $q_\mathrm{th}$ (yellow star) and for a wide range of momenta around  this value it remains very close to the continuum threshold, thus signifying a 
strong damping in the propagation.

\begin{figure}[ht]
\centering
\includegraphics[width=0.495\textwidth]{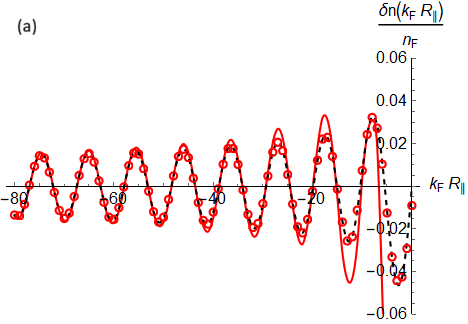}
\includegraphics[width=0.495\textwidth]{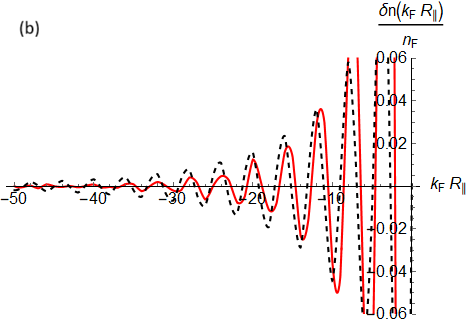}
\caption{Density modulation for $\tilde{V}_0=6$ and $v=1.25c_0$ shown as a red solid line for two different values of the interaction range, (a) $k_Fr_0=1.0$ and (b)  $k_Fr_0=0.5$. The black dashed line represents the zero-sound contribution given by Eq.\til\eqref{equ:dnanal}. 
In panel (a) the zero-sound contribution stemming from the numerical evaluation of Eq.\til\eqref{equ:dn2} for momenta $0 \leq q_\perp \leq q_\perp^c$, is also reported as red circles.}
\label{drhokFr02}
\end{figure}

However the existence of a non-vanishing length scale $r_0$ associated with the range of the interaction potential (shown as a long-dashed line for $k_Fr_0=1$ in Fig.\til\ref{dispkFr0V0qc}(a)) can significantly reduce the damping of the mode if $r_0 \gg 1/q_\mathrm{th}$. The threshold momentum  $q_\mathrm{th}$ is uniquely determined by the strength of the potential, as shown in Fig.\til\ref{dispkFr0V0qc}(b). 
Therefore for larger coupling strengths the damping of the mode at large distances (far-field) is expected to take place only for proportionally smaller interaction ranges $k_Fr_0$. Numerically we find that  if  $ r_0 \lesssim 2/q_\mathrm{th}$  the long range behavior of the density modulation becomes strongly suppressed, as displayed in panel (c) of Fig.\til\ref{drhokFr0} where $k_Fr_0 =0.8 <  2/q_\mathrm{th}=1.0$ at $\tilde{V}_0=4$ (see also Fig.\til\ref{dispkFr0V0qc}(b)).
We verify this empirical prediction by computing the density modulation for a different interaction strength $\tilde{V}_0=6$ (and $v=1.25c_0$), which yields $2/q_\mathrm{th}=0.625$. 
In Fig.\til\ref{drhokFr02} density modulations for two values of $k_Fr_0$, slightly below and slightly above $k_Fr_0 = 0.625$, are shown in order to illustrate the persistence of zero sound at large distances for $k_Fr_0=1.0$ and its strong decay for $k_Fr_0=0.5$.

\revise{
\section{Possible routes to experimental realization}
\label{sec:exprel}

The creation of a density modulation by means of a moving impurity in an ultracold atomic gas has interested a number of experiments regarding the onset of dissipation in both Bose and Fermi superfluids\til\cite{Carusotto-2006,Carusotto-2013,Xhani-2022,Xhani-2023,Xhani-2025,Xhani2-2025}. The Landau critical velocity associated to the relative motion between a superfluid and a disturbance \cite{SPUNTARELLI-2010,Pisani-2024}  represents the threshold above which viscosity appears owing to density fluctuations (in the form of sound or vortices) hence disrupting superfluidity.

In the present study the critical velocity above which a coherent density modulation can propagate is the zero sound speed. Given the current progress in the manipulation of the trapping geometry and of impurities in ultracold atomic gases\til\cite{Xhani-2022,Xhani2-2025}, one can envisage an annular trap  whereby the fermionic fluid is put into motion relative to a localized impurity, which in turn is geneated with a focused laser beam by a micro-mirror device.
Alternatively a dynamically movable impurity can be implemented by a piezo-actuated mirror
mount or an acousto-optical deflector as suggested in the study of quantum wakes in  ultra-cold Fermi gases in optical lattices\til\cite{Kolomeisky-2021}.
However, current state-of-the-art developments on ultracold repulsive Fermi gases does not allow to simulate strong interactions and the coherent excitation of the zero sound becomes severely hindered by its hybridization with the incoherent spectrum of {\it p-h} excitations. 

A promising system where the effect of  damping is expected to be reduced, even in the regime of weak interactions, is the two-dimensional dipolar Fermi gas\til\cite{Shlyapnikov-2013}.  The combination of the aforementioned technical advancements on trapping geometry and impurity creation or manipulation, with the detection of spatial density profiles via absorption imaging\til\cite{Carusotto-2006} appears as a realistic path to observing the zero sound density modulation. 

It is finally worth mentioning the detection of wakes formed behind charged objects in supersonic plasma flows\til\cite{Kolomeisky-2018,Miloch-2010} or laser-driven plasma wake fields\til\cite{PlasmaWake-1988} in relation to a recent experiment\til\cite{Fleury-2024} where a moving laser beam was driven onto a thin black polymer plate and found to excite elastic waves on this plate. As the excitation of zero sound triggers a solid-like response of a Fermi fluid and it is in itself an elastic wave, it is then possible to envisage a similar response on the surface of a film of liquid $^3$He and to view it by employing the same imaging technique used for visualization of turbulent flow
 in the normal component of superfluid $^4$He\til\cite{McKinsey-2010}.
}

\section{Conclusion}
\label{sec:conc}

In this work, we have investigated the density modulation generated by an  impurity moving through an interacting Fermi gas at zero temperature, thus weakly perturbing it. We have found that if the interaction of the gas is large with respect to its Fermi energy (as one would find in $^3$He) and if the impurity speed is above that of zero sound, the latter is excited predominantly over the incoherent background of {\it p-h} pairs. This effect causes the density modulation to develop behind the impurity and to be of a very long-range nature. In contrast, when the speed is below that of zero sound, only a localized perturbance is found. We have also examined our results against the modification of the strength \revise{and range of the gas interaction and of its functional form, noticing that they are all crucial} for the onset of the Landau damping of the zero sound mode.
We consider the present investigation to be of a preliminary nature. Recalling the specific features of the density excitation spectrum of $^3$He\til\cite{Glyde-2000,Faak-1994,Monaco-2007,Aldrich-1978} we have not taken into account the effective mass of the fermions, which works in favor of the separation between the zero sound dispersion and the continuum threshold hence reducing the Landau damping, neither we have considered the attractive part of the $^3$He pseudo-potential\til\cite{Aldrich-1978,Krotscheck-2010} giving rise to a roton-like excitation spectrum\til\cite{Pines-1981,Krotscheck-2010,Kolomeisky-2022}. \revise{Moreover, we considered only the effect of single {\it p-h} excitations\til\cite{Krotscheck-2011}}.

All the numerical data necessary to reproduce all figures in this work are available online\til\cite{zenodozs}.

\section*{Acknowledgements}

We acknowledge the use of the parallel computing cluster of the Open Physics Hub at the Department of Physics and Astronomy of the University of Bologna.

\paragraph{Funding information}

L.P. acknowledges financial support from the Italian Ministry of University and Research (MUR) under project PRIN2022, Contract No.~2022523NA7.

\bibliographystyle{unsrt}
\bibliography{bibliography}

@article{Kelvin-1887,
author = {Thomson, William },
title = {On the waves produced by a single impulse in water of any depth, or in a dispersive medium},
journal = {Proceedings of the Royal Society of London},
volume = {42},
number = {251-257},
pages = {80},
year = {1887},
doi = {10.1098/rspl.1887.0017},
URL = {https://royalsocietypublishing.org/doi/abs/10.1098/rspl.1887.0017},
eprint = {https://royalsocietypublishing.org/doi/pdf/10.1098/rspl.1887.0017}
}

@book{LL-1984,
title={{Electrodynamics of Continuous Media }},
author={L. D. Landau and E. M. Lifshitz},
publisher={Pergamon, Oxford},
year={1984}
}

@article{Carusotto-2006,
  title = {{Bogoliubov-\ifmmode \check{C}\else \v{C}\fi{}erenkov Radiation in a Bose-Einstein Condensate Flowing against an Obstacle}},
  author = {Carusotto, I. and Hu, S. X. and Collins, L. A. and Smerzi, A.},
  journal = {Physical Review Letters},
  volume = {97},
  issue = {26},
  pages = {260403},
  numpages = {4},
  year = {2006},
  publisher = {American Physical Society},
  doi = {10.1103/PhysRevLett.97.260403},
  url = {https://link.aps.org/doi/10.1103/PhysRevLett.97.260403}
}

@inbook{Carusotto-2013,
   title={{The Cerenkov Effect Revisited: From Swimming Ducks to Zero Modes in Gravitational Analogues}},
   ISBN={9783319002668},
   ISSN={1616-6361},
   url={http://dx.doi.org/10.1007/978-3-319-00266-8_6},
   DOI={10.1007/978-3-319-00266-8_6},
   booktitle={Analogue Gravity Phenomenology},
   publisher={Springer International Publishing},
   author={Carusotto, I. and Rousseaux, G.},
   year={2013},
   pages={109–144} }

@article{Kolomeisky-2018,
  title = {{Kelvin-Mach Wake in a Two-Dimensional Fermi Sea}},
  author = {Kolomeisky, E. B. and Straley, J. P.},
  journal = {Physical Review Letters},
  volume = {120},
  issue = {22},
  pages = {226801},
  numpages = {5},
  year = {2018},
  publisher = {American Physical Society},
  doi = {10.1103/PhysRevLett.120.226801},
  url = {https://link.aps.org/doi/10.1103/PhysRevLett.120.226801}
}

@incollection{Landau-1987,
title = {{Chapter IX - Shock Waves}},
booktitle = {Fluid Mechanics (Second Edition)},
publisher = {Pergamon},
edition = {Second},
pages = {313},
year = {1987},
isbn = {978-0-08-033933-7},
doi = {https://doi.org/10.1016/B978-0-08-033933-7.50017-9},
url = {https://www.sciencedirect.com/science/article/pii/B9780080339337500179},
author = {L.D. Landau and E.M. Lifshitz}
}

@article{Pines-1981,
    author = {Pines, D.},
    title = {{Elementary Excitations in Quantum Liquids}},
    journal = {Physics Today},
    volume = {34},
    number = {11},
    pages = {106},
    year = {1981},
    issn = {0031-9228},
    doi = {10.1063/1.2914350},
    url = {https://doi.org/10.1063/1.2914350},
    eprint = {https://pubs.aip.org/physicstoday/article-pdf/34/11/106/8288665/106_1_online.pdf},
}

@article{Monaco-2007,
  title = {{Zero Sound Mode in Normal Liquid       $^{3}\mathrm{He}$}},
  author = {Albergamo, F. and Verbeni, R. and Huotari, S. and Vank\'o, G. and Monaco, G.},
  journal = {Physical Review Letters},
  volume = {99},
  issue = {20},
  pages = {205301},
  numpages = {4},
  year = {2007},
  publisher = {American Physical Society},
  doi = {10.1103/PhysRevLett.99.205301},
  url = {https://link.aps.org/doi/10.1103/PhysRevLett.99.205301}
}

@article{Faak-1994,
  title={{Spin Fluctuations and Zero-Sound in Normal Liquid $^3\mathrm{He}$ studied by Neutron Scattering}},
  author={F{\aa}k, B. and Guckelsberger, K. and Scherm, R. and Stunault, A.},
  journal={Journal of Low Temperature Physics},
  volume={97},
  number={5},
  pages={445},
  year={1994},
  publisher={Springer}
}

@article{Navon-2025,
  title = {{Emergence of Sound in a Tunable Fermi Fluid}},
  author = {Huang, S. and Ji, Y. and Repplinger, T. and Assump\c{c}$\tilde{\mbox{a}}$o, G. G. T. and Chen, J. and Schumacher, G. L. and Vivanco, F- J. and Kurkjian, H. and Navon, N.},
  journal = {Physical Review X},
  volume = {15},
  issue = {1},
  pages = {011074},
  numpages = {13},
  year = {2025},
  publisher = {American Physical Society},
  doi = {10.1103/PhysRevX.15.011074},
  url = {https://link.aps.org/doi/10.1103/PhysRevX.15.011074}
}

@book{FW-2003,
title={{Quantum Theory of Many-Particle Systems}},
author={A. Fetter and J. D. Walecka},
publisher={Dover Publications},
year={2003}
}

@book{PN-1966,
title={{The Theory of Quantum Liquids}},
author={D. Pines and P. Nozieres},
publisher={W. A. Benjamin, New York},
year={1966}
}

@book{GV-2005,
title={{Quantum Theory of the Electron
Liquid}},
author={G. Giuliani and G. Vignale},
publisher={Cambridge University Press, Cambridge},
year={2005}
}

@book{Dobbs-2000,
title={{Helium Three}},
author={E. R. Dobbs},
publisher={Oxford University Press, New
York},
year={2000}
}

@incollection{Landau-1957,
title = {{Oscillations in a Fermi Liquid}},
editor = {D. {TER HAAR}},
booktitle = {Collected Papers of L.D. Landau},
publisher = {Pergamon},
pages = {731},
year = {1965},
isbn = {978-0-08-010586-4},
doi = {https://doi.org/10.1016/B978-0-08-010586-4.50096-1},
url = {https://www.sciencedirect.com/science/article/pii/B9780080105864500961},
}

@article{Abel-1966,
  title = {{Propagation of Zero Sound in Liquid ${\mathrm{He}}^{3}$ at Low Temperatures}},
  author = {Abel, W. R. and Anderson, A. C. and Wheatley, J. C.},
  journal = {Physical Review Letters},
  volume = {17},
  issue = {2},
  pages = {74--78},
  numpages = {0},
  year = {1966},
  month = {Jul},
  publisher = {American Physical Society},
  doi = {10.1103/PhysRevLett.17.74},
  url = {https://link.aps.org/doi/10.1103/PhysRevLett.17.74}
}

@article{Pines-1976,
  title = {Zero Sound and Spin Fluctuations in Liquid Helium-3},
  author = {Aldrich, C. H. and Pethick, C. J. and Pines, D.},
  journal = {Physical Review Letters},
  volume = {37},
  issue = {13},
  pages = {845--848},
  numpages = {0},
  year = {1976},
  month = {Sep},
  publisher = {American Physical Society},
  doi = {10.1103/PhysRevLett.37.845},
  url = {https://link.aps.org/doi/10.1103/PhysRevLett.37.845}
}

@article{Aldrich-1978,
title = {{Polarization Potentials and Elementary Excitations in Liquid $^3$He}},
journal = {Journal of Low Temperature Physics},
volume = {32},
number = {2},
pages = {689},
year = {1978},
issn = {0003-4916},
doi = {10.1007/BF00056653},
url = {https://doi.org/10.1007/BF00056653},
author = {Aldrich, C. H. and Pines, D.}
}

@article{Nettleton-1976,
  title={{Transport in the low-temperature limit of a Landau Fermi liquid}},
  author={Nettleton, R.E.},
  journal={Journal of Low Temperature Physics},
  volume={22},
  number={3},
  pages={407},
  year={1976},
  publisher={Springer}
}

@article{Rudnick-1980,
  title={{Zero sound and the viscoelasticity of liquid $^3$He}},
  author={Rudnick, I.},
  journal={Journal of Low Temperature Physics},
  volume={40},
  number={3},
  pages={287},
  year={1980},
  publisher={Springer}
}

@article{Scherm-1987,
  title = {Pressure dependence of elementary excitations in normal liquid helium-3},
  author = {Scherm, R. and Guckelsberger, K. and Fak, B. and Sk\"old, K. and Dianoux, A. J. and Godfrin, H. and Stirling, W. G.},
  journal = {Physical Review Letters},
  volume = {59},
  issue = {2},
  pages = {217},
  numpages = {0},
  year = {1987},
  publisher = {American Physical Society},
  doi = {10.1103/PhysRevLett.59.217},
  url = {https://link.aps.org/doi/10.1103/PhysRevLett.59.217}
}

@article{Vignale-1999,
  title = {{Elasticity of an electron liquid}},
  author = {Conti, S. and Vignale, G.},
  journal = {Physical Review B},
  volume = {60},
  issue = {11},
  pages = {7966},
  numpages = {0},
  year = {1999},
  publisher = {American Physical Society},
  doi = {10.1103/PhysRevB.60.7966},
  url = {https://link.aps.org/doi/10.1103/PhysRevB.60.7966}
}

@article{Glyde-2000,
  title = {{Effective Mass, Spin Fluctuations, and Zero Sound in Liquid ${}^{3}\mathrm{He}$}},
  author = {Glyde, H. R. and F\aa{}k, B. and Dijk, N. H. van and Godfrin, H. and Guckelsberger, K. and Scherm, R.},
  journal = {Physical Review B},
  volume = {61},
  issue = {2},
  pages = {1421},
  numpages = {0},
  year = {2000},
  publisher = {American Physical Society},
  doi = {10.1103/PhysRevB.61.1421},
  url = {https://link.aps.org/doi/10.1103/PhysRevB.61.1421}
}

@article{Schmets-2008,
  title = {Comment on ``Zero Sound Mode in Normal Liquid $^{3}\mathrm{He}$''},
  author = {Schmets, A. J. M. and Montfrooij, W.},
  journal = {Physical Review Letters},
  volume = {100},
  issue = {23},
  pages = {239601},
  numpages = {1},
  year = {2008},
  publisher = {American Physical Society},
  doi = {10.1103/PhysRevLett.100.239601},
  url = {https://link.aps.org/doi/10.1103/PhysRevLett.100.239601}
}

@article{Krotscheck-2010,
  title = {{Dynamic Many-Body Theory: Dynamics of Strongly Correlated Fermi fluids}},
  author = {B\"ohm, H. M. and Holler, R. and Krotscheck, E. and Panholzer, M.},
  journal = {Physical Review B},
  volume = {82},
  issue = {22},
  pages = {224505},
  numpages = {28},
  year = {2010},
  publisher = {American Physical Society},
  doi = {10.1103/PhysRevB.82.224505},
  url = {https://link.aps.org/doi/10.1103/PhysRevB.82.224505}
}

@article{Albergamo-2008,
  title = {Albergamo et al. Reply:},
  author = {Albergamo, F. and Verbeni, R. and Huotari, S. and Vank\'o, G. and Monaco, G.},
  journal = {Physical Review Letters},
  volume = {100},
  issue = {23},
  pages = {239602},
  numpages = {1},
  year = {2008},
  publisher = {American Physical Society},
  doi = {10.1103/PhysRevLett.100.239602},
  url = {https://link.aps.org/doi/10.1103/PhysRevLett.100.239602}
}

@article{Krotscheck-2011,
  title={Theoretical analysis of neutron and X-ray scattering data on 3He},
  author={Krotscheck, E. and Panholzer, M.},
  journal={Journal of Low Temperature Physics},
  volume={163},
  number={1},
  pages={1},
  year={2011},
  publisher={Springer}
}

@article{FETTER-1973,
title = {{Electrodynamics of a Layered Electron Gas. I. Single Layer}},
journal = {Annals of Physics},
volume = {81},
number = {2},
pages = {367},
year = {1973},
issn = {0003-4916},
doi = {https://doi.org/10.1016/0003-4916(73)90161-9},
url = {https://www.sciencedirect.com/science/article/pii/0003491673901619},
author = {A. L. Fetter}
}

@article{Kolomeisky-2022,
  title = {{Negative Group Velocity and Kelvin-like Wake Pattern}},
  author = {Kolomeisky, E. B. and Colen, J. and Straley, J. P.},
  journal = {Physical Review B},
  volume = {105},
  issue = {5},
  pages = {054509},
  numpages = {5},
  year = {2022},
  publisher = {American Physical Society},
  doi = {10.1103/PhysRevB.105.054509},
  url = {https://link.aps.org/doi/10.1103/PhysRevB.105.054509}
}

@manual{Mathematica,
  title        = {{Mathematica, Version 12.0}},
  author       = {{Wolfram Research, Inc.}},
  organization = {Wolfram Research, Inc.},
  address      = {Champaign, Illinois, USA},
  year         = {2019},
  note         = {\url{https://www.wolfram.com/mathematica/}}
}

@manual{zenodozs,
title = {{Data for "Density Modulation of Zero Sound"}},
author = {Pisani, Leonardo},
address ={{Zenodo Repository}}, 
year = {2025},
doi = {10.5281/zenodo.17238470},
note = {\url{https://zenodo.org/records/18248601}}
}

@article{Bohm-1951,
  title = {{A Collective Description of Electron Interactions. I. Magnetic Interactions}},
  author = {Bohm, D. and Pines, D.},
  journal = {Physical Review},
  volume = {82},
  issue = {5},
  pages = {625},
  numpages = {0},
  year = {1951},
  publisher = {American Physical Society},
  doi = {10.1103/PhysRev.82.625},
  url = {https://link.aps.org/doi/10.1103/PhysRev.82.625}
}

@article{Ye-2023,
  title = {{Zero-sound modes for the nuclear equation of state at supra-normal densities}},
  author = {Ye, J. and Margueron, J. and Li, N. and Jiang, W. Z.},
  journal = {Physical Review C},
  volume = {108},
  issue = {4},
  pages = {044312},
  numpages = {9},
  year = {2023},
  month = {Oct},
  publisher = {American Physical Society},
  doi = {10.1103/PhysRevC.108.044312},
  url = {https://link.aps.org/doi/10.1103/PhysRevC.108.044312}
}

@article{Giorgini-2008,
  title = {{Theory of ultracold atomic Fermi gases}},
  author = {Giorgini, S. and Pitaevskii, L. P. and Stringari, S.},
  journal = {Review of Modern Physics.},
  volume = {80},
  issue = {4},
  pages = {1215},
  numpages = {0},
  year = {2008},
  publisher = {American Physical Society},
  doi = {10.1103/RevModPhys.80.1215},
  url = {https://link.aps.org/doi/10.1103/RevModPhys.80.1215}
}

@article{Bloch-2008,
  title = {{Many-body physics with ultracold gases}},
  author = {Bloch, I. and Dalibard, J. and Zwerger, W.},
  journal = {Review of Modern Physics.},
  volume = {80},
  issue = {3},
  pages = {885},
  numpages = {0},
  year = {2008},
  publisher = {American Physical Society},
  doi = {10.1103/RevModPhys.80.885},
  url = {https://link.aps.org/doi/10.1103/RevModPhys.80.885}
}

@article{Bloch-2012,
  title={Quantum simulations with ultracold quantum gases},
  author={Bloch, I. and Dalibard, J. and Nascimbene, S.},
  journal={Nature Physics},
  volume={8},
  number={4},
  pages={267},
  year={2012},
  publisher={Nature Publishing Group}
}

@article{Chin-2010,
  title = {{Feshbach resonances in ultracold gases}},
  author = {Chin, C. and Grimm, R. and Julienne, P. and Tiesinga, E.},
  journal = {Review of Modern Physics.},
  volume = {82},
  issue = {2},
  pages = {1225},
  numpages = {0},
  year = {2010},
  publisher = {American Physical Society},
  doi = {10.1103/RevModPhys.82.1225},
  url = {https://link.aps.org/doi/10.1103/RevModPhys.82.1225}
}

@article{Baroni-2024,
  title={Quantum mixtures of ultracold gases of neutral atoms},
  author={Baroni, C. and Lamporesi, G. and Zaccanti, M.},
  journal={Nature Reviews Physics},
  pages={1},
  year={2024},
  publisher={Nature Publishing Group}
}

@article{Gross-2017,
  title={{Quantum simulations with ultracold atoms in optical lattices}},
  author={Gross, C. and Bloch, I.},
  journal={Science},
  volume={357},
  number={6355},
  pages={995},
  year={2017},
  publisher={American Association for the Advancement of Science}
}

@article{Landig-2015,
  title={Measuring the dynamic structure factor of a quantum gas undergoing a structural phase transition},
  author={Landig, R- and Brennecke, F. and Mottl, R. and Donner, T. and Esslinger, T.},
  journal={Nature Communications},
  volume={6},
  number={1},
  pages={7046},
  year={2015},
  publisher={Nature Publishing Group UK London}
}

@article{Torma-2016,
doi = {10.1088/0031-8949/91/4/043006},
url = {https://doi.org/10.1088/0031-8949/91/4/043006},
year = {2016},
month = {mar},
publisher = {IOP Publishing},
volume = {91},
number = {4},
pages = {043006},
author = {Törmä, P.},
title = {{Physics of ultracold Fermi gases revealed by spectroscopies}},
journal = {Physica Scripta},
}

@article{Hoinka-2012,
  title = {{Dynamic Spin Response of a Strongly Interacting Fermi Gas}},
  author = {Hoinka, S. and Lingham, M. and Delehaye, M. and Vale, C. J.},
  journal = {Physical Review Letters},
  volume = {109},
  issue = {5},
  pages = {050403},
  numpages = {5},
  year = {2012},
  publisher = {American Physical Society},
  doi = {10.1103/PhysRevLett.109.050403},
  url = {https://link.aps.org/doi/10.1103/PhysRevLett.109.050403}
}

@article{Hoinka-2013,
  title = {{Precise Determination of the Structure Factor and Contact in a Unitary Fermi Gas}},
  author = {Hoinka, S. and Lingham, M. and Fenech, K. and Hu, H. and Vale, C. J. and Drut, J. E. and Gandolfi, S.},
  journal = {Physical Review Letters},
  volume = {110},
  issue = {5},
  pages = {055305},
  numpages = {5},
  year = {2013},
  publisher = {American Physical Society},
  doi = {10.1103/PhysRevLett.110.055305},
  url = {https://link.aps.org/doi/10.1103/PhysRevLett.110.055305}
}

@article{Shlyapnikov-2013,
  title = {{Zero sound in a two-dimensional dipolar Fermi gas}},
  author = {Lu, Z.-K. and Matveenko, S. I. and Shlyapnikov, G. V.},
  journal = {Physical Review A},
  volume = {88},
  issue = {3},
  pages = {033625},
  numpages = {14},
  year = {2013},
  publisher = {American Physical Society},
  doi = {10.1103/PhysRevA.88.033625},
  url = {https://link.aps.org/doi/10.1103/PhysRevA.88.033625}
}

@article{SPUNTARELLI-2010,
title = {{Solution of the Bogoliubov–de Gennes equations at zero temperature throughout the BCS–BEC crossover: Josephson and related effects}},
journal = {Physics Reports},
volume = {488},
number = {4},
pages = {111},
year = {2010},
issn = {0370-1573},
doi = {https://doi.org/10.1016/j.physrep.2009.12.005},
url = {https://www.sciencedirect.com/science/article/pii/S0370157309002890},
author = {A. Spuntarelli and P. Pieri and G.C. Strinati}
}

@article{Pisani-2024,
  title = {Critical current throughout the BCS-BEC crossover with the inclusion of pairing fluctuations},
  author = {Pisani, L. and Piselli, V. and Strinati, G. Calvanese},
  journal = {Physical Review A},
  volume = {109},
  issue = {3},
  pages = {033306},
  numpages = {17},
  year = {2024},
  publisher = {American Physical Society},
  doi = {10.1103/PhysRevA.109.033306},
  url = {https://link.aps.org/doi/10.1103/PhysRevA.109.033306}
}

@article{Xhani-2025,
  title = {{Stability of persistent currents in superfluid fermionic rings}},
  author = {Xhani, K. and Barresi, A. and Tylutki, M. and Wlaz\l{}owski, G. and Magierski, P.},
  journal = {Physical Review Research},
  volume = {7},
  issue = {1},
  pages = {013225},
  numpages = {9},
  year = {2025},
  publisher = {American Physical Society},
  doi = {10.1103/PhysRevResearch.7.013225},
  url = {https://link.aps.org/doi/10.1103/PhysRevResearch.7.013225}
}

@misc{Xhani2-2025,
      title={{Tuning the Critical Current in Toroidal Superfluids via Controllable Impurities}}, 
      author={K. Xhani and G. Del Pace and N. Grani and D. Hernández-Rajkov and B. Donelli and G. Roati and L. Pezzè},
      year={2025},
      eprint={2511.10493},
      archivePrefix={arXiv},
      primaryClass={cond-mat.quant-gas},
      url={https://arxiv.org/abs/2511.10493}, 
}

@Article{Xhani-2023,
AUTHOR = {Xhani, K. and Del Pace, G. and Scazza, F. and Roati, G.},
TITLE = {{Decay of Persistent Currents in Annular Atomic Superfluids}},
JOURNAL = {Atoms},
VOLUME = {11},
YEAR = {2023},
NUMBER = {8},
ARTICLE-NUMBER = {109},
URL = {https://www.mdpi.com/2218-2004/11/8/109},
ISSN = {2218-2004},
DOI = {10.3390/atoms11080109}
}

@article{Xhani-2022,
  title = {{Imprinting Persistent Currents in Tunable Fermionic Rings}},
  author = {Del Pace, G. and Xhani, K. and Muzi Falconi, A. and Fedrizzi, M. and Grani, N. and Hernandez Rajkov, D. and Inguscio, M. and Scazza, F. and Kwon, W. J. and Roati, G.},
  journal = {Physical Review X},
  volume = {12},
  issue = {4},
  pages = {041037},
  numpages = {16},
  year = {2022},
  publisher = {American Physical Society},
  doi = {10.1103/PhysRevX.12.041037},
  url = {https://link.aps.org/doi/10.1103/PhysRevX.12.041037}
}

@article{Kolomeisky-2021,
  title = {{Quantum wakes in lattice fermions}},
  author = {Wampler, M. and Schauss, P. and Kolomeisky, E. B. and Klich, I.},
  journal = {Physical Review Research},
  volume = {3},
  issue = {3},
  pages = {033112},
  numpages = {14},
  year = {2021},
  publisher = {American Physical Society},
  doi = {10.1103/PhysRevResearch.3.033112},
  url = {https://link.aps.org/doi/10.1103/PhysRevResearch.3.033112}
}

@article{PlasmaWake-1988,
  title = {{Experimental Observation of Plasma Wake-Field Acceleration}},
  author = {Rosenzweig, J. B. and Cline, D. B. and Cole, B. and Figueroa, H. and Gai, W. and Konecny, R. and Norem, J. and Schoessow, P. and Simpson, J.},
  journal = {Physical Review Letters},
  volume = {61},
  issue = {1},
  pages = {98},
  numpages = {0},
  year = {1988},
  publisher = {American Physical Society},
  doi = {10.1103/PhysRevLett.61.98},
  url = {https://link.aps.org/doi/10.1103/PhysRevLett.61.98}
}

@ARTICLE{Du-Liu-2024,
  author={Du, H. and Liu, C.},
  journal={IEEE Transactions on Antennas and Propagation}, 
  title={An Improved FDTD Method to Calculate Nonlocal Response in Plasmonics}, 
  year={2024},
  volume={72},
  number={3},
  pages={2592-2599},
  doi={10.1109/TAP.2024.3363445}}

@article{Teperik-2013,
author = {T. V. Teperik and P. Nordlander and J. Aizpurua and A. G. Borisov},
journal = {Optics Express},
number = {22},
pages = {27306--27325},
publisher = {Optica Publishing Group},
title = {Quantum effects and nonlocality in strongly coupled plasmonic nanowire dimers},
volume = {21},
year = {2013},
url = {https://opg.optica.org/oe/abstract.cfm?URI=oe-21-22-27306},
doi = {10.1364/OE.21.027306}
}

@article{Miloch-2010,
doi = {10.1088/0741-3335/52/12/124004},
url = {https://doi.org/10.1088/0741-3335/52/12/124004},
year = {2010},
month = {nov},
publisher = {},
volume = {52},
number = {12},
pages = {124004},
author = {Miloch, Wojciech J},
title = {Wake effects and Mach cones behind objects},
journal = {Plasma Physics and Controlled Fusion}
}

@article{Fleury-2024,
  title = {Experimental observation of parabolic wakes in thin plates},
  author = {Rus, Janez and Bossart, Aleksi and Apffel, Benjamin and Mall\'ejac, Matthieu and Fleury, Romain},
  journal = {Physical Review Research},
  volume = {6},
  issue = {3},
  pages = {L032027},
  numpages = {7},
  year = {2024},
  publisher = {American Physical Society},
  doi = {10.1103/PhysRevResearch.6.L032027},
  url = {https://link.aps.org/doi/10.1103/PhysRevResearch.6.L032027}
}

@article{McKinsey-2010,
  title = {Visualization Study of Counterflow in Superfluid $^{4}\mathrm{He}$ using Metastable Helium Molecules},
  author = {Guo, W. and Cahn, S. B. and Nikkel, J. A. and Vinen, W. F. and McKinsey, D. N.},
  journal = {Physical Review Letters},
  volume = {105},
  issue = {4},
  pages = {045301},
  numpages = {4},
  year = {2010},
  publisher = {American Physical Society},
  doi = {10.1103/PhysRevLett.105.045301},
  url = {https://link.aps.org/doi/10.1103/PhysRevLett.105.045301}
}

\end{document}